\documentclass[3p,times,procedia,sort&compress]{elsarticle}
\usepackage{nupha_ecrc}
\usepackage[pdftex]{hyperref}
\usepackage[utf8]{inputenc}
\usepackage{multirow}

\volume{956}

\firstpage{1}

\journalname{Nuclear Physics A}

\runauth{C.\ Loizides}

\jid{nupha}

\jnltitlelogo{Nuclear Physics A}

\usepackage{amssymb}
\usepackage{amsthm}
\usepackage{lineno}
\usepackage[figuresright]{rotating}
\newif\ifcomment
\newif\ifdetails
\newif\ifarxiv
\arxivtrue
\newif\ifshortbib
\ifarxiv
\detailstrue
\else
\shortbibtrue
\fi

\newcommand{\snn}         {\ensuremath{\sqrt{s_{\rm NN}}}}
\newcommand{\pT}          {\ensuremath{p_{\rm T}}}
\newcommand{\kT}          {\ensuremath{k_{\rm T}}}
\newcommand{\Nch}         {\ensuremath{N_{\rm ch}}}
\newcommand{\Ncoll}       {\ensuremath{N_{\rm coll}}}
\newcommand{\Npart}       {\ensuremath{N_{\rm part}}}

\newcommand{\dAu}         {dAu}

\newcommand{\lsim}        {\,{\buildrel < \over {_\sim}}\,}
\newcommand{\gsim}        {\,{\buildrel > \over {_\sim}}\,}
\newcommand{\co}[1]       {\relax}
\newcommand{\nl}          {\newline}
\newcommand{\el}          {\\\hline\\[-0.4cm]}
\newcommand{\Ref}[1]      {Ref.~\cite{#1}}
\newcommand{\Sec}[1]      {Sec.~\ref{#1}}
\newcommand{\Fig}[1]      {Fig.~\ref{#1}}

\begin{document}
\begin{frontmatter}
\dochead{}

\title{Experimental overview on small collision systems at the LHC}
\author{Constantin Loizides}
\address{Lawrence Berkeley National Laboratory, Berkeley, California, USA}

\begin{abstract}
These conferences proceedings summarize the experimental findings obtained in small collision systems at the LHC,
as presented in the special session on ``QGP in small systems?'' at the Quark Matter 2015 conference.
\ifdetails (The arXiv version is significantly longer than the printed proceedings, with more details and a short discussion.)\fi
\end{abstract}

\begin{keyword}
heavy-ion, collectivity, high multiplicity, QGP
\end{keyword}
\end{frontmatter}


\section{Introduction}
\label{sec:intro}
Key evidence for the formation of a hot Quark-Gluon Plasma~(QGP) in nucleus--nucleus~(AA) collisions at high collision energies is the presence of jet quenching~\cite{Wiedemann:2009sh} and bulk collective effects~\cite{Voloshin:2008dg}, together with their absence in proton--nucleus~(pA) or deuteron--gold~(dAu) control experiments~\cite{Gyulassy:2004vg}.
The control measurements are needed to characterize the extent to which initial-state effects can be differentiated from effects due to final-state interactions in the QGP~\cite{Salgado:2011wc}. 
It is assumed that initial-state physics can be isolated since the QGP, if at all produced in such light--heavy ion reactions, is expected to be tenuous.

Indeed, in the case of hard processes at mid-rapidity, control experiments, both at RHIC in dAu collisions at $\snn=200$~GeV\cite{Back:2003ns,Adler:2003ii,Adams:2003im,Arsene:2003yk}, and at the LHC in pPb collisions at $\snn=5.02$ TeV~\cite{ALICE:2012mj,Abelev:2014dsa,Khachatryan:2015xaa,Adam:2015hoa,Khachatryan:2016xdg,Abelev:2013yxa,Aaij:2013zxa,Adam:2015iga,Aad:2015ddl,Abelev:2014hha,Abelev:2014oea,Aaij:2014mza,Khachatryan:2015uja,Khachatryan:2015sva,Adam:2015qda,Chatrchyan:2013exa}, demonstrated the absence of strong final state effects. 
In particular, the minimum bias pPb data can be well described by superimposing $N_{\rm coll}=A\sigma_{\rm pp}/\sigma_{\rm pPb}\approx 7$ independent pp collisions with only small modifications induced by the initial state or cold nuclear matter.

However, measurements of multi-particle correlations over large pseudorapidity~($\eta$) range in high multiplicity pPb events~\cite{CMS:2012qk,Abelev:2012ola,Aad:2012gla,Aad:2013fja,Chatrchyan:2013nka} exhibit remarkable similarities with results related to collective effects from  PbPb collisions at $\snn=2.76$~TeV~\cite{Loizides:2013nka}.\footnotemark[1]
In fact, soon after the start of the LHC program in 2010, a pronounced longitudinal structure in the two-dimensional angular correlation function measured for particle pairs over azimuthal angle and $\eta$ differences, $\Delta\varphi$ and $\Delta\eta$, with transverse momentum~($\pT$) of $1$--$3$~GeV/$c$ was observed in high-multiplicity pp events at 7 TeV~\cite{Khachatryan:2010gv}. 
The appearance of these ``ridge-like'' structures in high-multiplicity pp and pA events\co{ with similar characteristics than in PbPb} caused speculation of similar physics being at work in these small collision systems.
In particular, there is an ongoing debate as to whether they are of common hydrodynamic origin~\cite{Bozek:2013uha}, or created already by strong correlations in the initial state from gluon saturation~\cite{Dusling:2013qoz}.

In these QM15 proceedings, following the presentation at the conference~\cite{clqmtalk}, a summary on experimental results from small collision systems at the LHC with an emphasis on bulk properties will be given. 
\ifdetails
\enlargethispage{1cm}
\else
\enlargethispage{0.5cm}
\fi
\footnotetext[1]{Similar effects had been observed already at RHIC in dAu collisions~\cite{Adams:2004bi}, but were ignored until the LHC data were reported. For a summary of the recent experimental results at RHIC, see \Ref{stankusqm15}.}

\begin{table}[t!fp]
\begin{center}
  \begin{tabular}{p{4.5cm}|p{2.75cm}|p{2.75cm}|p{2.7cm}|\ifdetails p{1.6cm} \else p{1.2cm} \fi}
    Observable or effect                 & PbPb                                          & pPb (at high mult.)                  & pp (at high mult.)                         & Refs.\\
    \hline
    \hline
    Low $\pT$ spectra (``radial flow'')  & yes                                             & yes                                & yes                            & \cite{Abelev:2012wca,Abelev:2013vea,Chatrchyan:2013eya,Chatrchyan:2012qb,Andrei:2014vaa,Abelev:2013haa} \el
    Intermed.\ $\pT$ (``recombination'') & yes                                             & yes                                & yes                            & \cite{Andrei:2014vaa,Abelev:2013xaa,Abelev:2013haa,Abelev:2014uua,Khachatryan:2016yru,Adam:2015jca,Adam:2016dau} \el
    Particle ratios                      & GC level                                        & GC level except $\Omega$           & GC level except $\Omega$       & \cite{Adam:2016emw,Adam:2016bpr,Adam:2015vsf,ABELEV:2013zaa} \\
    Statistical model                    & $\gamma^{\rm GC}_s=1$, 10--30\%                   & $\gamma^{\rm GC}_s\approx1$, 20--40\%  & $\gamma^{\rm C}_s<1$, 20--40\% \footnotemark[2] & \cite{Floris:2014pta} \el
    HBT radii ($R(\kT)$, $R(\sqrt[3]{\Nch})$)        & $R_{\rm out}/R_{\rm side}\approx1$ \footnotemark[3]& $R_{\rm out}/R_{\rm side}\lsim1$         & $R_{\rm out}/R_{\rm side}\lsim1$       & \cite{Adam:2015vna,Adam:2015vja,Abelev:2014pja,Adam:2015pya,Aamodt:2011kd,CMS:2014mla,ATLAS:2015-054} \el
    Azimuthal anisotropy ($v_n$)\nl 
    (from two part.\ correlations)                   & $v_1-v_7$                                 & $v_1-v_5$                                  & $v_2$, $v_3$                  & \cite{CMS:2012qk,Abelev:2012ola,Aad:2012gla}\nl\cite{Aamodt:2011by,Chatrchyan:2011eka,Chatrchyan:2012wg,ATLAS:2012at,Aad:2014lta,Aad:2015gqa,CMS:2015zpa,Khachatryan:2016txc} \el
    Characteristic mass dependence                   & $v_2$, $v_3$ \footnotemark[4]             & $v_2$, $v_3$                               & $v_2$                         & \cite{Abelev:2014pua,Abelev:2012di,Adam:2016nfo,Khachatryan:2014jra,ABELEV:2013wsa,CMS:2015kua,Khachatryan:2016txc} \el
    Directed flow (from spectators)                  & yes                                       & no                                        & no                            & \cite{Abelev:2013cva}\el
    Higher order cumulants \nl(mainly $v_2\{n\}$, $n\ge4$)  & \mbox{``$4\approx6\approx8\approx$ LYZ''} \mbox{+higher harmonics} & \mbox{``$4\approx6\approx8\approx$ LYZ''} \mbox{+higher harmonics} & \mbox{``$4\approx6$''} \footnotemark[5] & \cite{Aad:2013fja,Chatrchyan:2013nka,Khachatryan:2016txc}\nl\cite{Aamodt:2010pa,ALICE:2011ab,Chatrchyan:2012ta,Abelev:2014mda,Chatrchyan:2013kba,Aad:2014vba,Khachatryan:2015waa,Adam:2016izf,CMS:2015ica} \el
    Weak $\eta$ dependence                           & yes                                       & yes                                       & not measured                  & \cite{Adam:2016ows,Aad:2014eoa,ATLAS:2011ah,Khachatryan:2016ibd,Adam:2015bka,Aaij:2015qcq,CMS:2015ica,Aaboud:2016jnr} \el
    Factorization breaking                           & yes ($n=2,3$)                             & yes ($n=2,3$)                             & not measured                  & \cite{Khachatryan:2015oea}\el
    Event-by-event $v_n$ distributions               & $n=2-4$                                   & not measured                              & not measured                  & \cite{Aad:2013xma} \el
    Event plane and $v_n$ correlations               & yes                                       & not measured                              & not measured                  & \cite{Aad:2014fla,Aad:2015lwa,ALICE:2016kpq} \el
\ifdetails
    Direct photons at low $\pT$                      & yes                                       & not measured                              & not measured \footnotemark[6] & \cite{Adam:2015lda}\el
    Jet quenching                                    & yes                                       & not observed \footnotemark[7]             & not measured \footnotemark[8] & \cite{Aad:2010bu,Aamodt:2010jd,Chatrchyan:2011sx,CMS:2012aa,Abelev:2012hxa,ALICE:2012ab,Aad:2014bxa,Adam:2015ewa,Aad:2015wga}\el
    Heavy flavor anisotropy                         & yes                                       & hint \footnotemark[9]                     & not measured                   & \cite{ALICE:2013xna,Abelev:2013lca,Abelev:2014ipa,Adam:2015pga}\el
    Quarkonia                                        & J/$\psi \uparrow$, $\Upsilon \downarrow$  & suppressed                                & not measured \footnotemark[8] & \cite{Abelev:2012rv,Adam:2015rba,Chatrchyan:2012lxa,Chatrchyan:2013nza,Abelev:2014zpa,Adam:2015jsa,Adam:2016ohd} \\
    \hline
    \hline
\fi
  \end{tabular}
  \caption{\label{tab:1}Summary of \ifdetails\else bulk \fi observables or effects (classified as known from AA data) in PbPb collisions, as well as in high multiplicity pPb and pp collisions at the LHC. References to key measurements for the various observables and systems are given. See text for details.}
\end{center}  
\end{table}
\footnotetext[2]{Statistical model results were extracted only for minimum-bias pp data.}
\footnotetext[3]{The dependence of the freeze-out parameters relative to the event-plane angle has also been measured~\cite{Loggins:2014uaa}.}
\footnotetext[4]{Identified particle $v_2\{4\}$ from cumulants has also been measured~\cite{Krzewicki:1609054}.}
\footnotetext[5]{Attempts to measure $v_2\{4\}$ in pp from cumulants before 2015 did not extract a real (non-imaginary) value for $v_2$~\cite{flowpp,CMS:2015kua}.}
\setcounter{footnote}{5}
\ifdetails
\footnotetext[6]{A measurement in minimum bias pp collisions at 7 TeV is consistent with zero at low $\pT$~\cite{Wilde:2012wc}.}
\footnotetext[7]{Jet quenching has not been directly observed~\cite{Adam:2014qja}, but in analogy with PbPb (where $v_2$ at high $\pT$ is thought to be caused by path-length dependent energy loss),
the possible non-zero $v_2$ at large $\pT$ may be indicative of parton energy loss in pPb~\cite{Aad:2014lta}.}
\footnotetext[8]{Conceptually, it is not straightforward to measure a modification because of difficulties to define a rigorous normalization for nuclear effects, which is already a challenge in the case of pA collisions due to the contribution from multiple-parton interactions~\cite{Adam:2014qja}.}
\footnotetext[9]{There are long-range two-particle correlation measurements of heavy-flavor electron--hadron correlations at mid-rapidity~\cite{Filho:2014vba} and inclusive muon--hadron correlations at forward rapidity~\cite{Adam:2015bka}, which may be indicative of non-zero $v_2$ for heavy-flavor particles.}
\setcounter{footnote}{9}
\fi

\section{Summary of observations in PbPb collisions}
\label{sec:sumpbpb}
In order to classify the observed phenomena in small systems, it is useful to briefly discuss the conclusions drawn from observations in PbPb collisions at the LHC, as compiled in Table~\ref{tab:1}.
Particle ratios and yields are described as a Grand Canonical ensemble by the statistical model with the strangeness under-saturation factor $\gamma_S\approx1$ and with only moderate deviations of 10--30\% from thermal yields~\cite{ABELEV:2013zaa,Floris:2014pta}.
Almost all particles (with $\pT\lsim2$ GeV/$c$) can be described by hydrodynamics, which efficiently translates initial spatial anisotropies~($\varepsilon_i$) into final momentum anisotropies~($v_i$) assuming that the pressure gradients have built up early in the evolution of the system~\cite{Ollitrault:1992bk}.
Hydrodynamic models quantitatively describe ``average observables'' as radial flow~\cite{Abelev:2012wca,Abelev:2013vea} and the transverse momentum ($\pT$) and pseudorapidity ($\eta$) dependence of the azimuthal anisotropies $v_n$~\cite{Aamodt:2011by,Chatrchyan:2011eka,Chatrchyan:2012wg,ATLAS:2012at,Aamodt:2010pa,ALICE:2011ab,Chatrchyan:2012ta,Adam:2016ows,Aad:2014eoa,ATLAS:2011ah}, as well as their characteristic mass dependence~\cite{Abelev:2014pua,Abelev:2012di,Adam:2016nfo,Khachatryan:2014jra}. 
Furthermore, hydrodynamic calculations can also (at least qualitatively) capture ``higher order'' details, such as the breaking of factorization due to event-plane angle decorrelations in $\pT$ and $\eta$~\cite{Khachatryan:2015oea}, $v_n$ distributions~\cite{Aad:2013xma}, as well as event-plane angle and event-by-event $v_n$ correlations across different harmonics~\cite{Aad:2014fla,Aad:2015lwa,ALICE:2016kpq}.
Approximately describing direct photon production at low $\pT$, such models determine the initial temperature of the QGP to exceed about 400~MeV in central collisions~\cite{Adam:2015lda}.
The freeze-out radii in 3 orthogonal directions~(``out'', ``side'', ``long'') can be deduced from measurements of quantum-statistics correlations between pairs of same-charge pions~(HBT) at low momentum transfer. 
The radii are found to scale with $\sqrt[3]{\Nch}$ (indicating a constant density at freeze-out) and to decrease with increasing pair momentum ($\kT$) as expected from hydrodynamics~\cite{Adam:2015vna}.
The size along the emission direction is similar to the geometric size of the system ($R_{\rm out}/R_{\rm side}\approx1$).
\ifdetails
Directed flow, both the rapidity-odd as well as the rapidity-even components, of charged particles at mid-rapidity was measured relative to the collision symmetry plane defined by the spectator nucleons, and evidence for dipole-like initial density fluctuations in the overlap region\co{, as well as weak correlation between fluctuating participant and spectator symmetry planes,} was found~\cite{Abelev:2013cva}.
\fi
At intermediate $\pT$~($2\lsim\pT\lsim5$ GeV/$c$), the yield of heavier particles is enhanced relative to that of lighter particles~\cite{Abelev:2013xaa,Abelev:2014uua,Khachatryan:2016yru}, an effect typically described by calculations employing a combination of hydrodynamics and quark coalescence~(or recombination)~\cite{Fries:2008hs}.
The created system is opaque for high-$\pT$ colored probes, which due to radiational and collisional energy loss (jet quenching) are strongly suppressed\ifdetails~\cite{Aad:2010bu,Aamodt:2010jd,Chatrchyan:2011sx,CMS:2012aa,Abelev:2012hxa,ALICE:2012ab,Aad:2014bxa,Adam:2015ewa,Aad:2015wga}\fi, and transparent for photons and other colorless probes, which roughly scale with $\Ncoll$~\ifdetails\cite{Chatrchyan:2011ua,Chatrchyan:2012vq,Chatrchyan:2012nt,Aad:2012ew,Chatrchyan:2014csa,Aad:2014bha,Aad:2015lcb}\else(see \cite{Roland:2014jsa,Armesto:2015ioy} for recent reviews)\fi.
Jet quenching leads to slightly modified jet fragmentation functions inside small jet cone sizes~($R=0.4$)\ifdetails~\cite{Chatrchyan:2014ava,Aad:2014wha}\fi, and most of the radiated energy appears at large angles~($R>0.8$)\ifdetails~\cite{Chatrchyan:2011sx,Khachatryan:2015lha}\fi.
Due to interactions and rescattering with the medium, even heavy-flavor particles exhibit finite anisotropies\ifdetails~\cite{ALICE:2013xna,Abelev:2013lca,Abelev:2014ipa,Adam:2015pga}\fi.
Strong support for the formation of a deconfined medium with color degrees of freedom is the relative enhancement of J/$\psi$ yields, in particular at low $\pT$, due to statistical recombination or regeneration\ifdetails~\cite{Abelev:2012rv,Adam:2015rba}\fi, and the sequential suppression of $\Upsilon$ states\ifdetails~\cite{Chatrchyan:2012lxa}\fi.

In summary, the system created in high-energy PbPb collisions at the LHC exhibits features expected for strongly-interacting partonic matter.
There are a few outstanding problems, such as the small p/$\pi$ ratio~\cite{Abelev:2012wca}\co{ (which probably is of dynamic origin)}, 
the non-zero elliptic flow measured for direct photons~\cite{Lohner:2012ct}, the rather large $v_3$ relative to $v_2$ measured in ultra-central PbPb collisions~\cite{CMS:2013bza}, 
and the significant suppression of three- and four-pion Bose-Einstein correlations compared to expectations from two-pion measurements~\cite{Adam:2015pbc}.
These problems will need to be resolved in the future, but are not expected to invalidate the general conclusions outlined above.

\section{Results from pPb collisions at high multiplicity}
\label{sec:ppb}
Table~\ref{tab:1} compares results from measurements in PbPb collisions, discussed in the previous section, with those in high multiplicity pPb collisions, where available.
As in PbPb collisions, the $\pT$ spectra of identified particles harden with increasing multiplicity~\cite{Chatrchyan:2013eya,Andrei:2014vaa,Abelev:2013haa} and, if interpreted in the same way as in PbPb collisions~(using a blast wave model fit) reveal, even larger radial flow than in PbPb events with similar multiplicity\ifdetails as predicted~\cite{Shuryak:2013ke}\fi.
With increasing multiplicity, particle ratios reach the grand canonical level except for the $\Omega$ particle, which however gets near~\cite{Adam:2016bpr,Adam:2015vsf}.
Nevertheless, at high multiplicity, the yields can be described by the statistical model with $\gamma_S\approx1$ and with deviations of 20--40\% (only sightly worse than in the case of PbPb) from the thermal expectation~\cite{Floris:2014pta}.
At first, long-range ($2<|\Delta\eta|<4$) ridge structures on the near side ($\Delta\varphi \approx 0$) were measured using two-particle angular correlations in high multiplicity pPb collisions~\cite{CMS:2012qk}, and found to have a much larger strength than that of similar long range structures in high-multiplicity pp collisions at 7 TeV~\cite{Khachatryan:2010gv}.
Later, azimuthal anisotropy coefficients $v_2$ and $v_3$ were extracted from per-trigger yields measured over azimuthal angle and $\eta$ differences, $\Delta\varphi$ and $\Delta\eta$, respectively~\cite{Abelev:2012ola,Aad:2012gla}.
The per-trigger yields obtained in low multiplicity events (which were shown to be similar to pp collisions at 2.76 and 7 TeV) were subtracted, in order to suppress the dominant background from jets.
Soon after, genuine four-particle correlations were also measured~\cite{Aad:2013fja,Chatrchyan:2013nka}, where the contribution from two-particle correlations were systematically removed~(referred to as the ``cumulant method''~\cite{Bilandzic:2010jr}).
The multiplicity and $\pT$ dependencies of $v_2$ and $v_3$ turned out to be similar to those obtained in PbPb collisions provoking speculations about the same physical origin in both systems~\cite{Loizides:2013nka}.
Recently, cumulant measurements have been extended to genuine six- and eight-particle correlations, as well as the Lee Yang Zero (LYZ) method, which is conceptually equivalent to measuring the correlation of all particles~\cite{Bhalerao:2003xf}.
As in PbPb collisions, it was found that, within $10$\%, $v_2\{4\}\approx v_2\{6\}\approx v_2\{8\} \approx v_2\{{\rm LYZ}\}$ in high multiplicity pPb collisions~\cite{Khachatryan:2015waa}\ifdetails, as shown in \Fig{fig:cms}\fi.
The presence of non-zero higher-order cumulants with similar magnitude, and in particular with a finite and similar value for $v_2\{{\rm LYZ}\}$ can be interpreted as crucial evidence for a hydrodynamically evolving system, where all particles, and hence the single-particle distribution itself, would be correlated to the initial-state symmetry planes.
\ifdetails
\begin{figure}[t!f]
\begin{center}
   \includegraphics[width=0.99\textwidth]{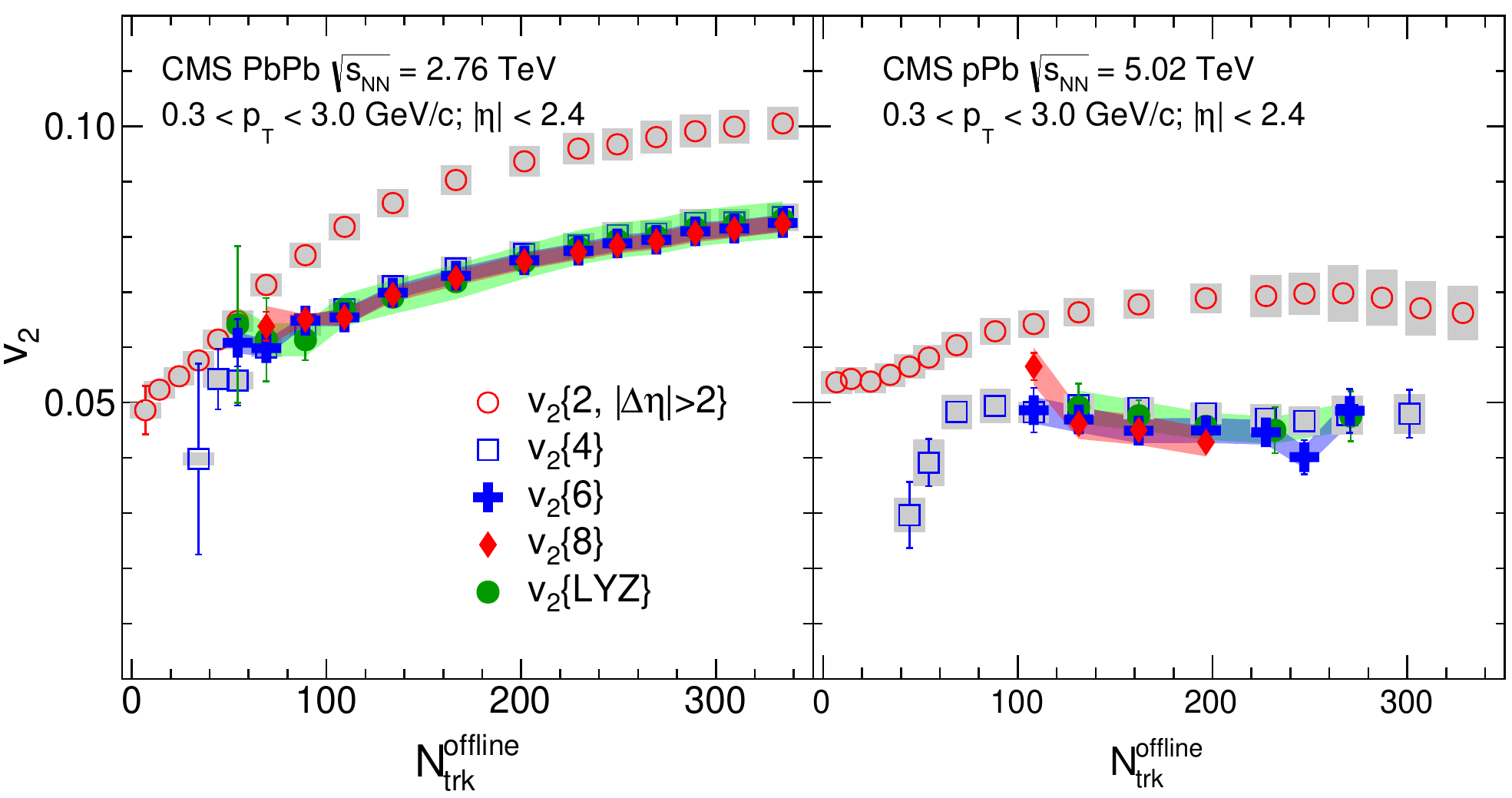}
   \caption{\label{fig:cms}The multiplicity dependence of $v_2\{n\}$ for $n=2,4,6,8$ and $v_2\{{\rm LYZ}\}$ averaged over $0.3<\pT<3$ GeV/$c$, for PbPb (left) and pPb (right panel) collisions.
            Figure taken from \Ref{Khachatryan:2015waa}.}
\end{center}
\end{figure}
\fi
\begin{figure}[t!f]
\begin{center}
   \includegraphics[width=0.48\textwidth]{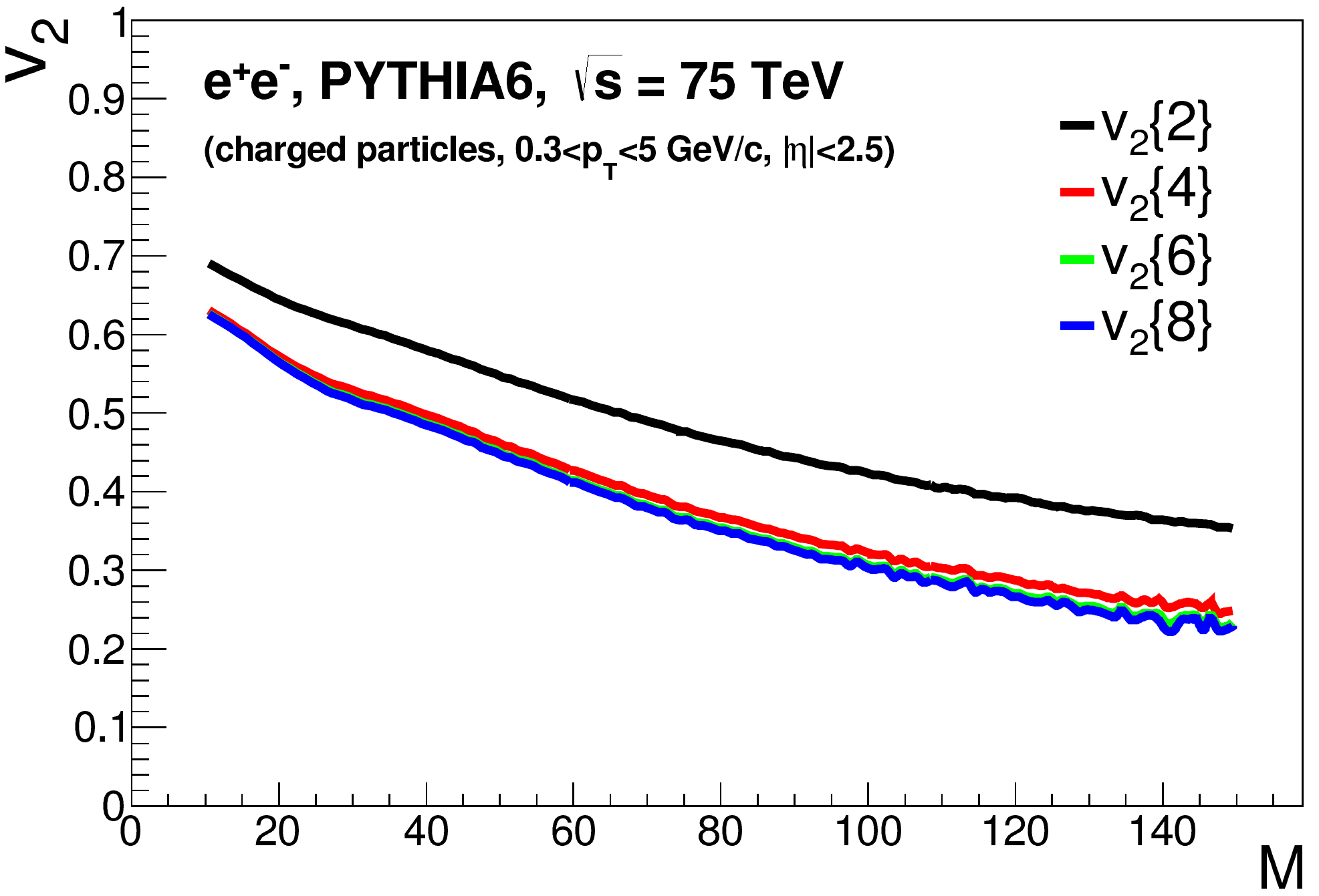}
   \hspace{0.25cm}
   \includegraphics[width=0.48\textwidth]{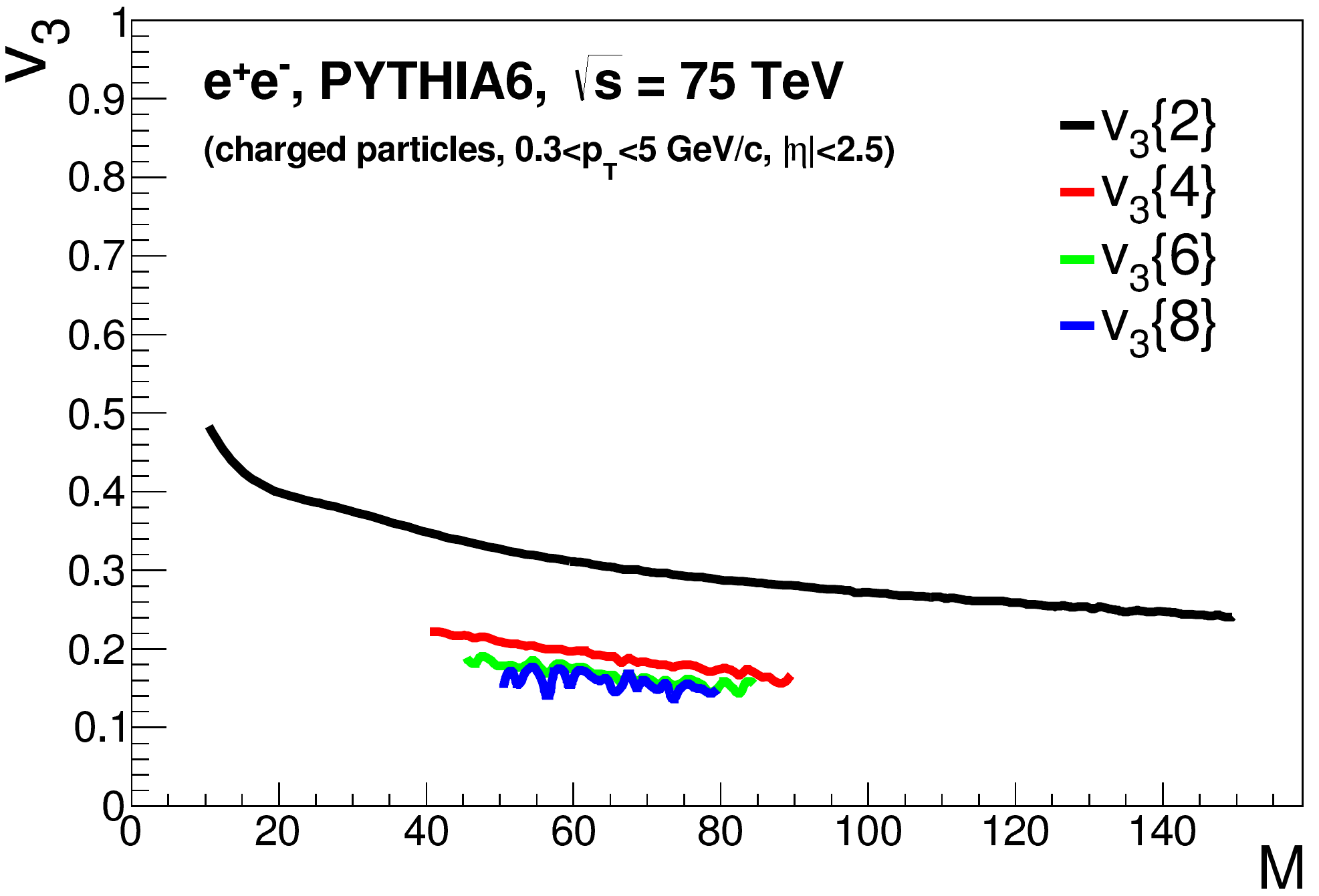}
   \caption{\label{fig:cumee}Second (left panel) and third (right panel) harmonics extracted from 2-, 4-, 6- and 8-particle cumulants vs charged particle multiplicity ($M$) in e$^+$e$^-$ collisions generated with PYTHIA (v6.425) at $\sqrt{s}=75$~TeV.
            Charged particles with $0.3<\pT<5$~GeV/$c$ and within $|\eta|<2.5$ are used in the calculation. For $v_3$, the higher-order results could only be reliably extracted in a limited range of $M$.}
\end{center}
\end{figure}
However, there are also arguments in disfavor of this interpretation.
Firstly, higher-order cumulants can have similar values even though the underlying correlations originate from a different mechanism than hydrodynamics. This is demonstrated in \Fig{fig:cumee} by comparing various multi-particle cumulants extracted for e$^+$e$^-$ collisions generated with the PYTHIA event generator~\cite{pythia6}.
\ifdetails
However, such apparent collectivity is a trivial consequence if the number of sources for particle production is just a few.
\fi
Secondly, in case of a narrow underlying distribution of $v_2$~(where the width is significantly smaller than the mean), the higher moments generally only differ by parametrically small amounts~\cite{Jia:2014pza}. 
Thirdly, models based on only quantum chromodynamics\ifdetails, and not involving hydrodynamics\fi, which do not require final-state interactions among quarks and gluons, have also been suggested to explain the observed multi-particle correlations\ifdetails, but have not provided any quantitative predictions yet\fi~\cite{Gyulassy:2014cfa,McLerran:2014uka}.
However, the $v_2$ and $v_3$ coefficients also exhibit a similar mass dependence as in PbPb~\cite{ABELEV:2013wsa,Khachatryan:2014jra}, with the $v_2$ and $v_3$ for heavier particle being depleted at low $\pT$, and the protons and $\Lambda$ crossing the $v_2$ (and $v_3$) of pions and neutral kaons, respectively, between $\pT$ of $1.5$--$3$ GeV/$c$.
In PbPb, the same effect is generally accepted to be a robust prediction of hydrodynamics, caused by the interplay of radial flow with the higher harmonics at lower $\pT$~\cite{Shen:2011eg} and recombination at higher $\pT$~\cite{Fries:2008hs}, affected also by the presence of a dense hadron gas at later stages of the collision evolution.
The breaking of factorization due to $\eta$ and $\pT$ dependent event-plane angle fluctuations is also found to be similar between pPb and PbPb collisions~\cite{Khachatryan:2015oea}.
As in PbPb, a rather weak $\eta$ dependence for $v_2$ and $v_3$ is measured~\cite{Khachatryan:2016ibd,Adam:2015bka,Aaij:2015qcq,CMS:2015ica}, in particular when the results are corrected for a possible effect arising from event-plane angle decorrelation along $\eta$.
Also, similarly to PbPb, HBT radii are found to scale with $\sqrt[3]{\Nch}$, to decrease with increasing $\kT$, and $R_{\rm out}/R_{\rm side}$ approaches $\approx 1$ at low $\kT$~\cite{Abelev:2014pja,Adam:2015pya,ATLAS:2015-054,CMS:2014mla}.
In the intermediate $\pT$ range~($2\lsim\pT\lsim5$ GeV/$c$), the spectra of identified particles show a characteristic enhancement in pPb compared to pp collisions~(scaled by $\Ncoll\approx7$)~\cite{Adam:2015jca,Adam:2016dau}, often referred to as ``Cronin'' enhancement. 
In the same $\pT$ region, particle ratios of p/$\pi$ and $\Lambda$/K$^0_{\rm S}$ show an enhancement also, whose magnitudes increase and maxima shift to higher $\pT$ with increasing multiplicity\co{, suggesting a connection with the Cronin enhancement}, depicting a similar $\pT$ dependence to PbPb collisions~\cite{Abelev:2013haa,Khachatryan:2016yru}.

The results discussed so far are reported in intervals of multiplicity measured either at mid- or forward rapidity, generally called ``event activity classes''.
To ensure that the conclusions are qualitatively robust against variation of the event selection, the measurements are sometimes reported for several different event activity definitions.
Quantitative interpretation of the data requires, however, that one takes into account that the observable and event activity may be correlated by the complex interplay between hard, multiple semi-hard and soft processes.
This correlation is suspected to induce strong changes of high-energy jet~\cite{ATLAS:2014cpa}, dijet~\cite{Chatrchyan:2014hqa}, D-meson~\cite{Adam:2016mkz} and $\Upsilon$ production~\cite{Chatrchyan:2013nza} with event activity. 
In order to compare data with calculations addressing nuclear effects\co{ in event activity classes}, event activity classes have also been related to an estimate of the number of collisions (or participants with $\Npart=\Ncoll+1$ in pPb) via a Glauber model~\cite{Adam:2014qja,Aad:2015zza}.
Due to the ordering of events according to their activity, pPb events with more than average activity per nucleon--nucleon collision populate the higher multiplicity classes, while with less than average activity the lower multiplicity classes~\cite{Abelev:2014mva}.
The event selection bias leads to a large variation of the nuclear modification factor, $Q_{\rm pPb}=1/\Ncoll \, Y_{\rm pPb}/Y_{\rm pp}$, and complicates a direct measurement of nuclear modification\co{ from charged particle spectra in high multiplicity pPb collisions}~\cite{Adam:2014qja}\ifdetails\footnotemark.\footnotetext{High-$\pT$ related results measured in minbias pPb collisions at mid-rapidity~\cite{ALICE:2012mj,Abelev:2014dsa,Khachatryan:2015xaa,Adam:2015hoa,Khachatryan:2016xdg,Abelev:2013yxa,Aaij:2013zxa,Adam:2015iga,Aad:2015ddl,Abelev:2014hha,Abelev:2014oea,Aaij:2014mza,Khachatryan:2015uja,Khachatryan:2015sva,Adam:2015qda,Chatrchyan:2013exa}, mentioned in \Sec{sec:intro}, do not have this problem. The charged particle $R_{\rm pPb}$ at high $\pT$ is consistent with unity. Initial discrepancies between results from ALICE, ATLAS and CMS, are now identified to be related to the interpolated pp reference~\cite{CMS:2015bfa}, and final confirmation can be expected from the analysis of pp collisions at $\sqrt{s}=5.02$~TeV taken end of 2015.}\else.\fi
Nevertheless, by comparing the data with a superposition of PYTHIA pp events, in which the event activity selection is performed in the same as in the data, it can be deduced that $Q_{\rm pPb}$ for charged particles, when integrated over $10$ to $20$ GeV/$c$, is not significantly altered by the nuclear environment. 
This conclusion is supported by a different approach, where events are ordered based on the zero-degree energy of slow neutrons produced in the collision on the Pb-going side\co{, which is expected to be related the least to a dynamical bias at mid-rapidity}, and an estimate of $\Ncoll$ from a data-driven approach inspired by the wounded nucleon model~\cite{Adam:2014qja}.
Furthermore, no significant nuclear modification has been seen in (charged) jet spectra~\cite{Adam:2016jfp}, D-meson yields~\cite{Adam:2016mkz} nor dijet $\kT$~\cite{Adam:2015xea} at mid-rapidity either.
While direct jet quenching~\cite{Shen:2016egw} has not been observed, possible non-zero $v_2$ values of about $5$\% at $\pT$ up to $10$ GeV/$c$ albeit with rather large uncertainty have been reported~\cite{Aad:2014lta}.
In analogy to PbPb (where $v_2$ at high $\pT$ is thought to be caused by path-length dependent energy loss), this observation may be a hint of parton energy loss in pPb.
However, the effect can also been qualitatively described by initial-state gluon correlations without (collective) final state interactions~\cite{Schenke:2015aqa}.
At forward rapidities, there are hints of final-state effects reported in the multiplicity dependence of J/$\psi$ $Q_{\rm pPb}$~\cite{Adam:2015jsa} and in the observed suppression of $\psi$(2S) relative to J/$\psi$~\cite{Abelev:2014zpa,Adam:2016ohd}. 
Lastly, there also are long-range two-particle correlation measurements of heavy-flavor electron--hadron correlations at mid-rapidity~\cite{Filho:2014vba} and inclusive muon--hadron correlations at forward rapidity~\cite{Adam:2015bka}, both revealing a double-ridge structure, which may be indicative of non-zero $v_2$ for heavy-flavor particles.

\section{Results from pp collisions at high multiplicity}
\label{sec:pp}
Table~\ref{tab:1} compares results from measurements in PbPb and pPb collisions, discussed in the previous sections, with those in high multiplicity pp collisions, where available.
Identified particle spectra at low $\pT$ show the same characteristic features as in pPb collisions, in particular a significant hardening with increasing multiplicity~\cite{Chatrchyan:2012qb,Andrei:2014vaa}.
Furthermore, at intermediate $\pT$~($2\lsim\pT\lsim5$ GeV/$c$) particle ratios of p/$\pi$ and $\Lambda$/K$^0_{\rm S}$ show a similar enhancement as in pPb collisions~\cite{Andrei:2014vaa,Khachatryan:2016yru,Adam:2016emw}. 
The data are consistent with the interpretation of strong radial flow developed in high multiplicity pp collisions~\cite{Kalaydzhyan:2015xba}.
Alternatively, however, microscopic effects such as the Color Reconnection~(CR) mechanism of strings implemented in PYTHIA can also qualitatively explain the data\ifdetails~\cite{Ortiz:2013yxa}\else~\cite{Ortiz:2013yxashort}\fi.
PYTHIA with CR describes the increase of the average $\pT$ with multiplicity~\cite{Abelev:2013bla} and the multiplicity dependence of charge-dependent correlations~(balance function)~\cite{Adam:2015gda} quite well.
Measured particle ratios increase with increasing multiplicity in the same way as in pPb collisions~\cite{Adam:2016emw}.
The statistical model has so far only been applied to minimum bias pp collisions and, when treated as a canonical ensemble, was found to describe the yields with $\gamma^{\rm C}_s<1$ and deviations of only about $20$--$40$\% from the expected yields~\cite{Floris:2014pta}.
HBT radii show the same trends with $\Nch$ and $\kT$ as in pPb collisions, and at high multiplicity $R_{\rm out}/R_{\rm side}$ approaches $\approx 1$ at low $\kT$~\cite{Abelev:2014pja,Aamodt:2011kd,CMS:2014mla}, consistent with the interpretation of collective flow~\cite{Hirono:2014dda}.

At the same multiplicity, near-side ridge yields extracted from long-range two-particle angular correlations at 13 TeV~\cite{Khachatryan:2015lva} agree with those measured at 7 TeV~\cite{Khachatryan:2010gv}, while they are a factor of 10 and 4 smaller than in PbPb and pPb collisions, respectively.
In pp even more so than in pPb collisions, jets are the dominant source of two-particle correlations, and in order to extract the long-range Fourier coefficients their contribution to the long-range correlations needs to be subtracted first.
Thus, as for the pPb results, the per-trigger yield measured at low multiplicity scaled by a factor $F$ is subtracted from that measured at high multiplicity, before the Fourier coefficients~($A_n$) from the long-range $\Delta\eta$ projection are computed.
Finally, the two-particle Fourier coefficients $V_n=A_n/B$ are obtained by normalizing the Fourier coefficients relative to the baseline~($B$) extracted in the high multiplicity events~(usually denoted with capital letters to distinguish them from those of the single-particle azimuthal distribution denoted with lower letters).
The single-particle Fourier coefficients are then $v_n=\sqrt{V_n}$\ifdetails\footnotemark\footnotetext{In case different $\pT$ ranges are used for trigger and associated particles, they are usually computed by relying on \mbox{factorization} $V_n(\pT^1,\pT^2)=v_n(\pT^1)v_n(\pT^2)$ , which is found to approximately hold by selecting different $\pT$ ranges for associated particles}\fi.
The factor $F$ can be obtained in various ways, for example as ratio of the short-range near-side yield in high multiplicitiy to that in low multiplicity events after the long-range $\Delta\eta$ contributions have been subtracted.
Assuming that the away-side jet shape per-trigger yields do not significantly change with multiplicity, the long-range jet contribution to the per-trigger yields should be significantly reduced.
In this way, significant ($\pT$ integrated) $v_2$ and $v_3$ coefficients have recently been reported in high multiplicity~($M$)\footnotemark pp collisions~\cite{Aad:2015gqa,CMS:2015kua}, with values of about 5\% and 1\%, respectively, about $30$\% and $50$\% smaller than in pPb at similar multiplicity. 
\footnotetext{Multiplicity is given as the number of reconstructed charged tracks with $\pT>0.4$~GeV/$c$ in $|\eta|<2.4$ for CMS and $\pT>0.5$~GeV/$c$ in $\eta<2.5$ for ATLAS. To roughly account for the tracking inefficiency scale by 1.2, and by 1.5 to extrapolate to $\pT=0$. Hence $M=100$ approximately corresponds to $5$--$6$ times the average multiplicity in pp, and is similar to the charged particle density at mid-rapidity of 70--80\% PbPb and 20--25\% CuCu collisions at $\snn=2.76$ TeV and $\snn=200$ GeV, respectively\ifdetails~\cite{Aamodt:2010cz,Alver:2010ck}\fi.}The
$\pT$ dependence of $v_2$ is found to be independent of $M$ above $M\gsim75$, and independent of collision energy (from 2.76 to 13 TeV).\footnotemark
\footnotetext{The integrated $v_2$ can not easily be compared between the ATLAS and CMS results as ATLAS uses $0.5<\pT<5$ while CMS $0.3<\pT<3$ GeV/$c$. However, 
within (quite large) systematic uncertainties the $\pT$-differential measurements of ATLAS and CMS agree, as shown in \cite{clqmtalk}.}Below 
$M\lsim70$, $v_2$ decreases with decreasing multiplicity.
Interestingly in this region, an ambiguity in the precise determination of $v_2$ arises~\cite{Aad:2015gqa} as the extracted $v_2$ values are found to depend on the number of tracks used to define the low-multiplicity event sample\ifdetails~(see right panel of \Fig{fig:atlas}, where the effect is shown in the case of ATLAS' template fit method further mentioned below). Larger $v_2$ are extracted with event classes, which are defined with even smaller multiplicity of only up to 10 and 5 tracks, respectively\fi.
\ifdetails
\begin{figure}[t!f]
\begin{center}
   \includegraphics[width=0.99\textwidth]{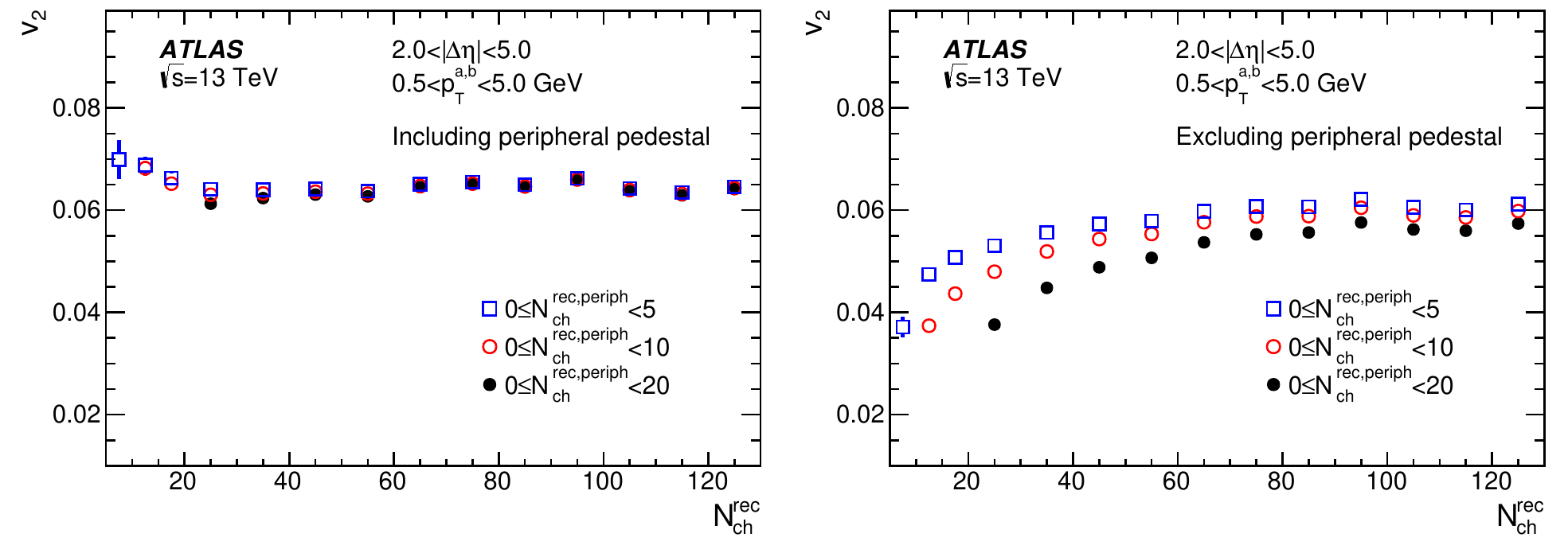}
   \caption{\label{fig:atlas}The multiplicity dependence $v_2$ for $0.5<\pT<5.0$ GeV/$c$ for three different low-multiplicity samples in pp collisoins at 13 TeV using the ATLAS template fit method allowing for non-zero $v_2$ in the low multiplicity sample (left panel) or assuming it to be zero (right panel), in which case the template fit is equivalent to the subtraction method. Minbias collisions roughly correspond to $N^{\rm rec}_{\rm ch}\approx 15$. Auxiliary figure 13 from \Ref{Aad:2015gqa}.}
\end{center}
\end{figure}
\fi
The cause for the dependence of $v_2$ on the choice of the low multiplicity event class is not yet clear.
It could originate from diffractive processes, or string breaking and resonance decay effects, which contribute to the correlation, since quite low $\pT$ particles are included in the measurement.
Both effects may significantly influence the shape of low multiplicity event class.\ifdetails\footnotemark\fi
Hence, it may be useful to repeat the analysis with a lower $\pT$ cut of at least $\pT>0.7$ GeV/$c$, for which near- and away-side associated yields have a similar slope with multiplicity~\cite{Abelev:2013sqa}.
In the case of CMS, contributions from lowest multiplicity are reduced, because events with less than 10 tracks are excluded from the low multiplicity sample, however, a lower $\pT$ cut of $0.3$ instead of $0.5$ GeV/$c$ is used.  
\ifdetails
\footnotetext{It is important to define the correlation function as a ``ratio of sums'' and not as a ``sum of ratios'', as briefly mentioned in \cite{Abelev:2012ola}, in order to avoid inherent multiplicity-dependent effects in the measurement of the correlation function.}Though, \else Though, \fi 
as mentioned in \Sec{sec:ppb}, one usually ensures that the low-multiplicity sample used in the subtraction procedure does not reveal unknown features~(for example in the case of pPb by comparing with minimum bias pp~\cite{Abelev:2012ola}). 
Since changing the range of the low multiplicity sample changes the measured $v_2$, one can attempt to correct the measured $v_2$ for an assumed fraction of $v^0_2$ present in the low multiplicity sample.
To achieve this a template fit of the scaled low multiplicity yield with an added $G(1+2V_2\cos(2\Delta\varphi))$ term is used by ATLAS to extract \ifdetails$\frac{GV_2-FG_0V_2^0}{G-FG_0}$, thus measuring approximately $V_2$ if $V_2^0\simeq V_2$\footnotemark ($G_0$ is the baseline in the low-multiplicity class)\else$V_2$\fi~\cite{Aad:2015gqa}.
\ifdetails\footnotetext{This way to extract $V_2$ is called ``Including peripheral pedestal'' by ATLAS to indicate that the resulting $V_2$ from the template fit is not further corrected by the baseline of the higher multiplicity event class. This is different from the case, which is labeled ``Excluding peripheral pedestal'', where the $V_2$ returned by the fit is divided by the baseline of the higher multiplicity class, as explained for the subtraction method above.}In \else In \fi
this case, the measured $v_2$ values are found to be rather constant with multiplicity for $M\gsim20$, essentially the same for different choices of constructing the low multiplicity sample, and about $10$\% larger than the results from the subtraction method at high $M$\ifdetails~(see left panel of \Fig{fig:atlas})\fi.
Below $M<20$ the $v_2$ values slightly increase with decreasing $M$ indicating that there are indeed larger modifications to the long-range correlation structures present in such low multiplicity events, perhaps related to diffraction and hadronization, as discussed above.
\ifdetails
The consistency of the results support the assumption of a constant $v_2$, which can not be proven for the lower multiplicity events, but holds already for $M>20$.
Instead, if one assumes that $v_2=0$ for $M<20$, then the measured $v_2$ values increase with multiplicity, reaching asymptotically similar values as for the constant $v_2$ case, because the assumption about $v_2$ in the low multiplicity event has ever decreasing influence. 
Above $M\gsim70$, the two methods lead to $v_2$ values with differences within about $20$\%, representing a robust measurement of $v_2$ at high multiplicity.
At lower $M$, the spread of the $v_2$ values indicates that more studies are needed\co{, such as a better characterization of the influence from diffraction and jet,}, but independent of the assumption even in events with $M>30$ (roughly twice of minimum bias collisions) a significant value of $v_2\approx0.04$ is obtained. 
\fi
Measurements involving four- and six-particle cumulants~\cite{Khachatryan:2016txc}, which earlier did not reach conclusive results~\cite{flowpp,CMS:2015zpa}, reveal that $v_{2}\{4\}\approx v_{2}\{6\}>0$ even in pp collisions at high multiplicity~($M>75$) consistent with the emergence of collective behavior.
However, these measurements, in particular for small $M$, need sufficiently high statistics and can not resolve the conceptual difference between including and excluding the pedestal in the subtraction discussed above.
Already now results of $v_2$ versus $\pT$ for K$^0_{\rm S}$ and $\Lambda$ have been presented~\cite{Khachatryan:2016txc} and for high multiplicity reveal a similar characteristic mass dependence as in pPb and PbPb collisions, which is absent in low multiplicity events\ifdetails, and may be interpreted as strong support for a hydrodynamic behavior~\cite{Torrieri:2013aqa}\fi.
\ifdetails
The shape and even the magnitude of the long-range component of two particle-pseudorapidity correlations are found to be similar for pp, pPb and PbPb at similar multiplicity~\cite{Aaboud:2016jnr}, which may be used to constrain calculations of the initial state for the three systems.
Other effects, in particular related to jet quenching\cite{Zakharov:2013gya}, are not straightforward to measure because of difficulties to define a rigorous normalization for nuclear effects, which is already a challenge in the case of pPb due to the contribution from multiple-parton interactions~\cite{Adam:2014qja}.
In pp even more than in pPb collisions there is a complex interplay between hard, multiple semi-hard and soft processes~\cite{Chatrchyan:2013nza,Abelev:2013sqa,ALICE:2011ac,Abelev:2012rz,LeonVargas:2012rw,Adam:2015ota,Aad:2015ziq}, which needs to be disentangled from nuclear modification.
\fi

\ifdetails
\section{Discussion and concluding remarks}
\else
\section{Concluding remarks}
\fi
\label{sec:conclusions}
The previous sections briefly described measurements, and their common interpretation, in PbPb, as well as in high multiplicity pPb and pp collisions, following the summary given in Table~\ref{tab:1}.
Weak collectivity (i.e.\ the agreement of higher-order cumulant and LYZ methods) is established within the experimental uncertainties in PbPb and pPb collisions, and starts to emerge even in pp collisions~($v_{2}\{4\}\approx v_{2}\{6\}$).
However, it should also be seen in e$^+$e$^-$ collisions~(see \Fig{fig:cumee}).
Most observables in PbPb and high multiplicity pPb collisions can even be explained assuming strong collectivity, i.e.\ as a system described by thermo- and hydrodynamics.
There is less information available at high multiplicity for pp collisions, however ---where available--- the trends are similar as for pPb\ifdetails~(for a recent review on these topics, see \Ref{Dusling:2015gta})\fi.
Naturally, similar observations could be caused by similar or even common physics. 
Assuming that the underlying physics is the same for the observed phenomena, what are the possible common explanations?
\ifdetails

i) A natural candidate for a system exhibiting strong collective effects is of course the sQGP~\cite{Gyulassy:2004zy}, which appears to flow almost like a perfect fluid with its shear-viscosity to entropy-density ratio $\eta/s$ close to the theoretical lower bound~\cite{Shen:2015msa}. 
In pPb and even more in pp collisions, one expects a stronger radial flow\footnotemark than in PbPb at similar multiplicities, because an approximately scale-invariant system would be smaller but hotter, leading to longer cooling and expansion~\cite{Shuryak:2013ke,Basar:2013hea}.
\footnotetext{Originally, radial flow was suggested to be present in pp collision even at ISR collision energies~\cite{Shuryak:1979ds}.}As 
for PbPb, hydrodynamic calculations quantitatively describe the small-system data~\cite{piotrqm}, and even the magnitude of $v_2$ in pp has been successfully predicted~\cite{Avsar:2010rf,CasalderreySolana:2009uk}.
In case of the proton, however, sub-nuclear scales in modelling the initial state  become important and, since they are not well known, result in a larger ambiguity of the calculations~\cite{Bzdak:2013zma}.
A challenge for the sQGP picture is the presence of a finite $v_2$ at $10$ GeV/$c$~(even though with large uncertainties)~\cite{Aad:2014lta} without direct evidence of jet quenching in the charged-particle $Q_{\rm pPb}$~\cite{Adam:2014qja}, while jet calculations indicate a significant, but with the present uncertainties difficult to measure, jet quenching effect~\cite{Shen:2016egw}.
In the low density limit, $v_2$ is expected to scale with multiplicity~(as a proxy for the density of the system), while $v_2$ should saturate in the hydrodynamic limit~\cite{Heiselberg:1998es}.
Taking the $v_2$ results from \mbox{ATLAS} in pp collisions at face value~\cite{Aad:2015gqa}, one would have hydrodynamically generated $v_2$ down to very low multiplicity, i.e.\ sQGP droplets form in essentially every (non-diffractive) pp event at 13 TeV, which seems quite surprising.\footnotemark\footnotetext{However, the formation of a deconfined system in such collisions can not apriori be excluded. The charged-particle mid-rapidity density at 13 TeV for events with one track (INEL$>$0) is $6.46\pm0.19$~\cite{Adam:2015pza}, which would lead to an energy density above $1.5$~GeV/fm$^3$ for $\tau<1$~fm/$c$ assuming an average $\pT$ as measured for 7 TeV pp collisions\cite{Abelev:2013bla}.} 
However, significant progress has been achieved in the past, by pushing the picture of the almost perfect fluid ever further, following ``naturalness'' arguments, exemplified in the following:
The large $v_2$ in central CuCu collisions at $\snn=200$ GeV, which was found to be indeed larger than in AuAu collisions and hence larger than expected from hydrodynamics, was explained\co{ in the hydrodynamic picture}
by postulating the relevance of eccentricity fluctuations, and in particular the ``participant eccentricity''~\cite{Alver:2006wh}.
The participant eccentricity, which predicted characteristic flow fluctuations consistent with a later measurement~\cite{Alver:2010rt}, was generalized to ``triangularity'' and highlighted the relevance of higher moments in the initial state\co{ based on analogy arguments}~\cite{Alver:2010gr}. 
The triangular structures only became directly visible in the PbPb data~\cite{ALICE:2011ab,Aamodt:2011by}, and today the pPb and pp data could not have been understood in the hydrodynamic picture without acknowledging the relevance of initial state fluctuations.
Further, prompted by the similarity of $v_3$ in pPb and PbPb~\cite{Chatrchyan:2013nka}, the hydrodynamic paradigm was successfully tested in $^3$HeAu collisions at RHIC~\cite{Adare:2015ctn}, where the use of Helium-3\co{ instead of deuteron} was expected to enhance the observed $v_3$ due to the intrinsic triangular shape of its wavefunction~\cite{Nagle:2013lja}\co{~(called ``geometric engineering'')}.
Nevertheless, despite its success, it must be emphasized that hydrodynamics can not explain how the system is brought into (approximate) equilibriium, and its use by construction hides any underlying microscopic physics picture. 

ii) A second possibility is to explain the observed phenomena with parton transport models, such as BAMPS~\cite{Xu:2004mz} or AMPT~\cite{Lin:2004en}, which employ non-equilibrium parton dynamics.
These calculations attempt to microscopically describe the underlying physics, and may be able to lead from weak to strong collectivity depending on the parton density and interaction cross sections.
Indeed, early calculations were argued to support the effective use of ideal hydrodynamics because one needed to inflate the (elastic) partonic cross sections ($\sigma$) from about $3$ to $45$ mb (i.e.\ to hadronic level) for a realistic initial gluon density\co{ of about ${\rm d}N/{\rm d}y$} of $1000$--$1500$ at mid-rapidity, in order to explain the early $v_2$ results in AuAu collisions at $\snn=130$~\cite{Molnar:2001ux}.
Today we know that AMPT with much lower cross sections of $1.5$ to $3$ mb can reproduce most of the experimental findings~\cite{Lin:2014tya,Ma:2014pva,Bzdak:2014dia,Ma:2016fve}.
AMPT, which was used to substantiate the postulation of triangularity~\cite{Alver:2010gr}, had even incorporated the underlying initial-state fluctuations and their influence on final azimuthal particle spectra before they were identified to be relevant.
However, in AMPT the azimuthal anisotropies are produced mainly by the anisotropic parton escape probability as a response to the initial spatial eccentricity with no or only a few parton interactions, not necessarily by pressure-driven collective flow. \footnotemark\footnotetext{A finite contribution to $v_2$ from an azimuthally non-uniform escape probability\co{ from the surface} is expected to be generated in a pre-equilibrium phase, because only colliding particles contribute to the eventual formation of approximate equilibrium, while those with no or few collisions leave the collision zone, contributing to escape $v_2$. It is an open question to what extent viscous hydrodynamic calculations include the effect, and whether it indeed is conceptually different from collective flow, or not a necessary ingredient. However, $v_2$ from escape does not require a hydrodynamic phase.} 
The contribution to the final $v_2$ from the escape probability is also found to be significant (about \mbox{$30$--$40$\%} even for $\sigma=45$ mb) in semi-central (about $20$\%) nucleus--nucleus collisions~\cite{He:2015hfa,Lin:2015ucn}.
Additionally, AMPT generates the characteristic dependence of $v_2$ on the particle mass from kinematics in the implemented coalescence approach and hadronic rescatterings~\cite{Li:2016flp}.
These recent conclusions drawn from studies using AMPT question aspects of the perfect liquid paradigm, perhaps even in semi-central and central AA collisions.
However, despite the fact that AMPT describes a wealth of the data, it is important to realize that the above conclusions depend on an interplay of questionable and not yet fully understood concepts such as string melting (resulting in the presence of a large number of quark/antiquark pairs and zero gluons), $\pT$ dependent parton formation and spatial coalescence.
Hence, follow-up studies, in particular on the relevance of the escape mechanism, using a different transport model, such as BAMPS, and viscous hydrodynamic calculations are needed.
Recently, it also was reported that non-hydrodynamic~(free streaming) evolution can generally create equal or larger radial flow than hydrodynamics (with $\eta/s=0.08$)~\cite{Romatschke:2015dha,paulqm}.
In pPb collisions free streaming leads to large $v_3$, but considerably smaller $v_2$ than from hydrodynamics, while as expected it cannot account for the observed harmonics in PbPb collisions.
Since pre-equilibrium dynamics and hadronic interactions are expected to play a significant role in small systems, a beam energy scan of \dAu\ collisions at RHIC may shed light on the question of whether equilibrium dynamics is indeed the source of the observed collectivity~\cite{Koop:2015trj}.

iii) The third possibility could be correlations present in the initial state followed by a suitable ``evolution'', perhaps leading from Yang-Mills evolution in smaller systems to hydrodynamic evolution in larger systems. 
The glasma graph framework derived in the context of the Color Glas Condensate~(CGC) is able to describe many features of two-particle correlations~\cite{Dusling:2013qoz,Dusling:2015rja,Schenke:2016lrs}, including the long range nature, the double ridge structure, the strength and shape, the non-trivial multiplicity dependence, as well as mass ordering of mean $\pT$ and $v_2$ in pp and pPb collisions.
However, the original calculation can neither account for genuine multi-particle correlations, nor for the presence of odd harmonics, most notably $v_3$.
Explaining these data has been achieved by using classical Yang-Mills simulations on the lattice~\cite{Schenke:2015aqa,Lappi:2015vta}, which incorporate contributions from glasma graphs\co{ as genuine non-factorizable contributions}, from perturbatively disconnected graphs (which connect through event-by-event breaking of rotational symmetry into domains with a size related to the inverse of the saturation scale~\cite{Dumitru:2014yza,Lappi:2015vha}), as well as from final state interactions generated by the Yang-Mills evolution.
In these simulations, the gluons already have a large $v_2$ in the initial state, albeit a $v_3$ of zero. 
However, after the time $\tau\le0.4$ fm/$c$ final state effects induced by the classical Yang-Mills evolution generate a non-zero $v_3$ with only little modification of $v_2$. 
In contrast, Yang-Mills simulations of PbPb collisions exhibit much smaller values of $v_2$ and $v_3$, since the contribution from disconnected graphs is significantly reduced due to the increase in the number of uncorrelated domains\co{the locally generated anisotropy due to the breaking of rotational invariance is depleted with the increase in the number of uncorrelated domains}.
Therefore, additional collective flow effects are needed to explain the PbPb data. 
However, so far these Yang-Mills simulations neglect contribution from jet graphs (which at higher order may become relevant) and effects from hadronization, which are also considered to be important~\cite{Esposito:2015yva}.
Furthermore, the forward and backward $v_2$ results in pPb~\cite{Adam:2015bka} question ideas based on the CGC, because the saturation scale dependence on rapidity would predict $v_2$ on the Pb side to be smaller than on the proton side, while the data indicate an opposite trend, qualitatively consistent with expectations from hydrodynamics and AMPT~\cite{Bozek:2013sda,Bozek:2015swa}.
In general, there is not yet a single framework that can quantitatively describe all data on similar footing, but effort in this direction is ongoing~\cite{Schlichting:2016xmj}.
\else
i)~Strongly-coupled QGP (sQGP), i.e.\ thermo- and hydrodynamics, maybe ``at the edge'', see~\Ref{piotrqm};
ii)~Non-equilibrium parton dynamics, maybe leading from weak to strong collectivity, see~\Ref{paulqm};
iii)~Initial-state (CGC) correlations followed by Yang-Mills or hydrodynamic ``evolution'', see~\Ref{Schlichting:2016xmj}.
\fi
In conclusion, the study of small systems may allow us\co{ i)} to probe dynamics rather than equilibrium\ifdetails~(with a further handle on possible viscous effects)\else\fi,\co{ ii)} to potentially validate or invalidate the ``perfect fluidity'' paradigm, and\co{ iii)} to test fundamental QCD due to relevance of sub-nucleon degrees of freedom in the initial state.

\ifdetails
\section*{Acknowledgments}
\else
\vspace{0.2cm}\noindent{\bf Acknowledgments}
\fi
I would like to thank the organizers for creating an interesting and stimulating conference. 
\ifdetails
Fruitful discussions with J.~Schukraft are acknowledged.
Special thanks to M.~Fasel, J.~Kamin and L.~Milano for critical reading of the draft.
\else
\enlargethispage{1cm}
\fi
This work is supported in part by the U.S. Department of Energy, Office of Science, Office of Nuclear Physics, under contract number DE-AC02-05CH11231.


\ifshortbib
\bibliographystyle{elsarticle-num-mod}
\else
\bibliographystyle{elsarticle-num}
\fi
\bibliography{clexp,clth,clproc}
\end{document}